# CrowdInside: Automatic Construction of Indoor Floorplans

Moustafa Alzantot
Wireless Research Center
Egypt-Japan Univ. of Sc. and Tech. (E-JUST)
Alexandria, Egypt
moustafa.alzantot@ejust.edu.eg

Moustafa Youssef
Wireless Research Center
Alexandria University and E-JUST
Alexandria, Egypt
moustafa.youssef@ejust.edu.eg

## ABSTRACT
The existence of a worldwide indoor floorplans database can lead to significant growth in location-based applications, especially for indoor environments. In this paper, we present *CrowdInside*: a crowdsourcing-based system for the automatic construction of buildings floorplans. *CrowdInside* leverages the smart phones sensors that are ubiquitously available with humans who use a building to automatically and transparently construct accurate motion traces. These accurate traces are generated based on a novel technique for reducing the errors in the inertial motion traces by using the points of interest in the indoor environment, such as elevators and stairs, for error resetting. The collected traces are then processed to detect the overall floorplan shape as well as higher level semantics such as detecting rooms and corridors shapes along with a variety of points of interest in the environment.

Implementation of the system in two testbeds, using different Android phones, shows that *CrowdInside* can detect the points of interest accurately with 0.2% false positive rate and 1.3% false negative rate. In addition, the proposed error resetting technique leads to more than 12 times enhancement in the median distance error compared to the state-of-the-art. Moreover, the detailed floorplan can be accurately estimated with a a relatively small number of traces. This number is amortized over the number of users of the building. We also discuss possible extensions to *CrowdInside* for inferring even higher level semantics about the discovered floorplans.

## 1. INTRODUCTION
During the last decade, there has been a rapid growth in location based applications, including location-enabled social networking, direction finding, and advertisement. This has been driven by the flourishing of smart phones and mobile devices, location determination technologies, and wireless Internet connectivity. A key requirement to many of these location-based applications is the availability of a map to display the user location on. This map can be a street map, in case of outdoor applications, or a floorplan, in case of indoor applications. Traditionally, outdoor location-based services providers, such as Google Maps, Bing Maps, FourSquare, etc, provide outdoor street maps for almost all regions around the globe. However, the indoor equivalent floorplans are currently very limited, affecting the ubiquity and spread of indoor location-based applications. Recently, a number of commercial systems for indoor direction finding have started to emerge, e.g. Point Inside and Micello Indoor Maps. In late 2011, Google Maps started to provide detailed floorplans for a few malls and airports in the U.S. and Japan. Nevertheless, all these systems depend on **manually** building the floor plan. Manual addition/editing of all buildings floorplans around the world requires an enormous cost and effort which may be unaffordable. In addition, keeping these floorplans up to date is another challenge.

In this paper, we introduce *CrowdInside* as a automatic floorplan construction system. *CrowdInside* leverages the ubiquity of smart phones to infer information about the building floorplan along with other semantic information. In particular, today's smart phones have an array of sensors, e.g. inertial sensors (accelerometers, compasses, and gyroscopes), that can be used to construct traces of movement in a transparent manner to the users. People walking in their homes, offices, and even visitors collect these traces and send them for processing by *CrowdInside*. Using this crowdsourcing approach, *CrowdInside* can provide the general layout of a building, identify the rooms and corridor locations and shapes, along with identifying other points of interest, such as elevators, stairs, and escalators.

*CrowdInside*, however, has to address a number of challenges including handling the smart phones noisy sensors, estimating the positions of points of interests in the building, detecting rooms and corridors shapes, and identifying doors locations.

Implementation of *CrowdInside* in a shopping mall and a university campus shows that it can estimate the floorplans with high accuracy with a relatively small number of traces. Such a system enables a wide set of indoor location based systems including, indoor directions finding, fine-grained location-based ads, indoor social networking applications, and ubiquitous indoor localization.

In summary, we provide the following contributions in this paper:

- We present the *CrowdInside* system architecture for leveraging the smart phones sensors in a crowdsourcing approach to automatically estimate the indoor floorplans for virtually any building around the globe.
- We provide techniques for estimating points of interests (or anchor points) in the environment (such as building entrances, elevators, stairs, and escalators) based on the phones inertial sensors with high accuracy.

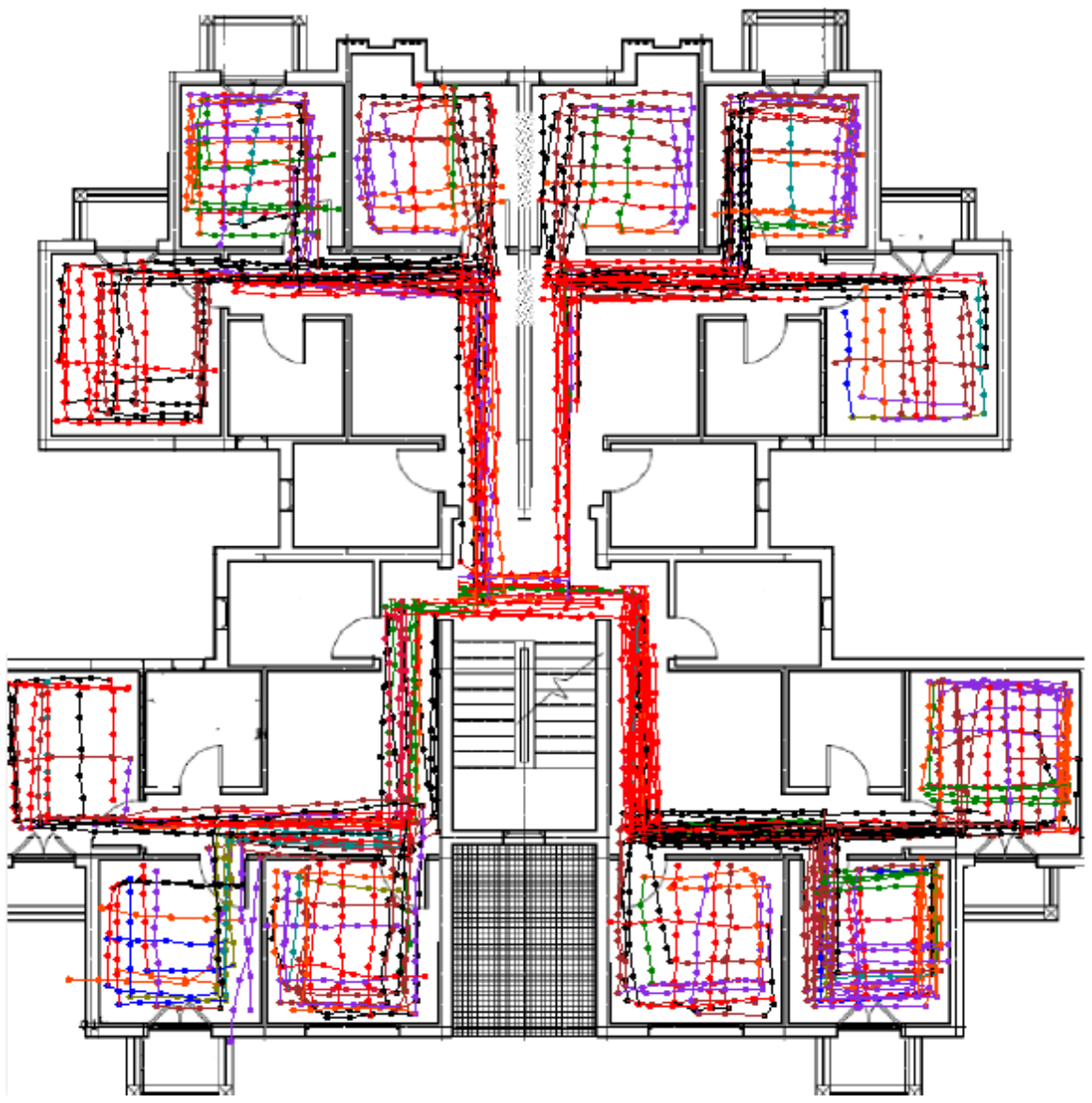

**Figure 1: Typical motion traces inside a building.**

- We provide a novel technique for constructing accurate indoor user traces based on the noisy inertial sensors in today's commodity smart phones. The proposed technique depends on resetting the accumulation of error by leveraging the detected anchor points.
- We employ classification techniques to separate corridors from rooms and further apply clustering techniques to separate the rooms from each other.
- We show how to identify the rooms shapes using computational geometry techniques.
- Finally, we implement the system on different Android phones (Samsung Nexus S, Nexus One, Galaxy Ace and Galaxy Tab) and evaluate it in a campus building and a mall.

The rest of this paper is organized as follows: We start by giving an overview of the *CrowdInside* system and how it can construct accurate traces in Section 2. Section 3 gives the details of the floorplan estimation module. We evaluate *CrowdInside* in Section 4. Section 5 discusses the related work. Finally, Section 6 concludes the paper and gives directions for future work.

## 2. SYSTEM DESIGN

Our system design is based on a crowdsourcing approach, where measurements from sensors embedded in mobile devices are collected from users moving naturally inside the buildings. The intuition behind this is that a large number of motion traces can provide an adequate description of the building's layout. Figure 1 shows an example for the motion traces collected from a number of users moving inside a building. As the number of traces increases, we get a better idea of the building layout. *CrowdInside* employs further processing to provide more semantic information, such as separating rooms and corridors, points of interest (such as elevators, stairs, escalators, etc).

Figure 2 shows our system architecture. The system consists of three main module: (a) the **Data Collection Module** is responsible for collecting measurements from users' devices, (b) the **Traces Generation Module** is responsible for building accurate motion traces based on a novel anchor-based error resetting technique and (c) the **Floorplan Estimation Module** that separates the corridors from the rooms and detects the rooms boundaries. We describe the details of the first two modules in this section and leave the details of the Floorplan Estimation Module to Section 3.

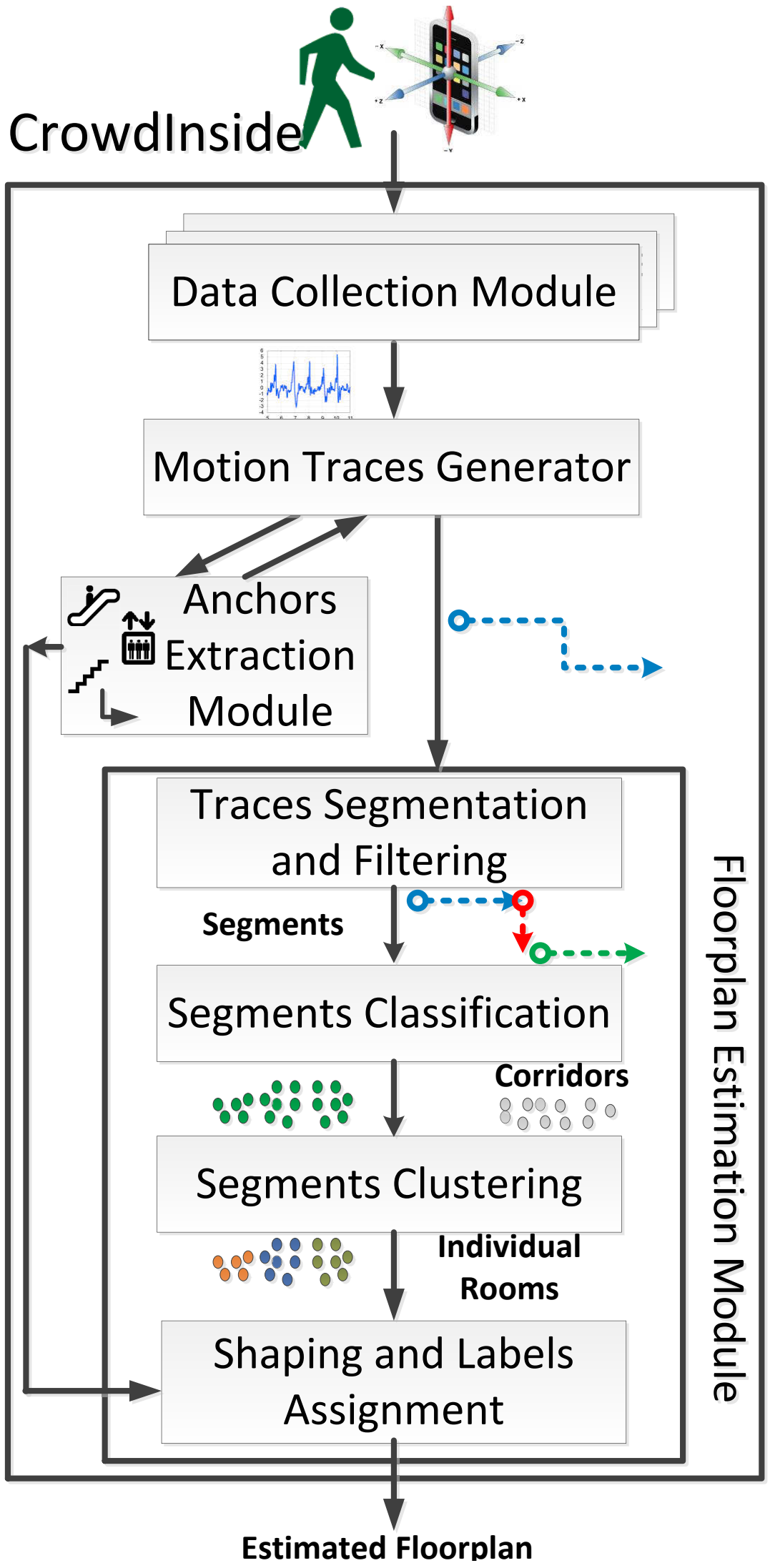

**Figure 2: *CrowdInside* system architecture.**

### 2.1 Data Collection Module

This module is responsible for collecting measurements from the various sensors embedded in the users' mobile devices. The timestamped measurements collected can be buffered and then sent opportunistically to server in the cloud for later processing when a connection is available to reduce the communications cost and/or save energy. Data collected are measurements from sensors including: accelerometers, magnetometers, gyroscopes, and the received WiFi signal strength values from available access points. The GPS is also queried with a low duty cycle to detect the user's transition from outdoors to indoors. The duty cycle can be set adaptively according to user's current position (e.g. more frequently when the

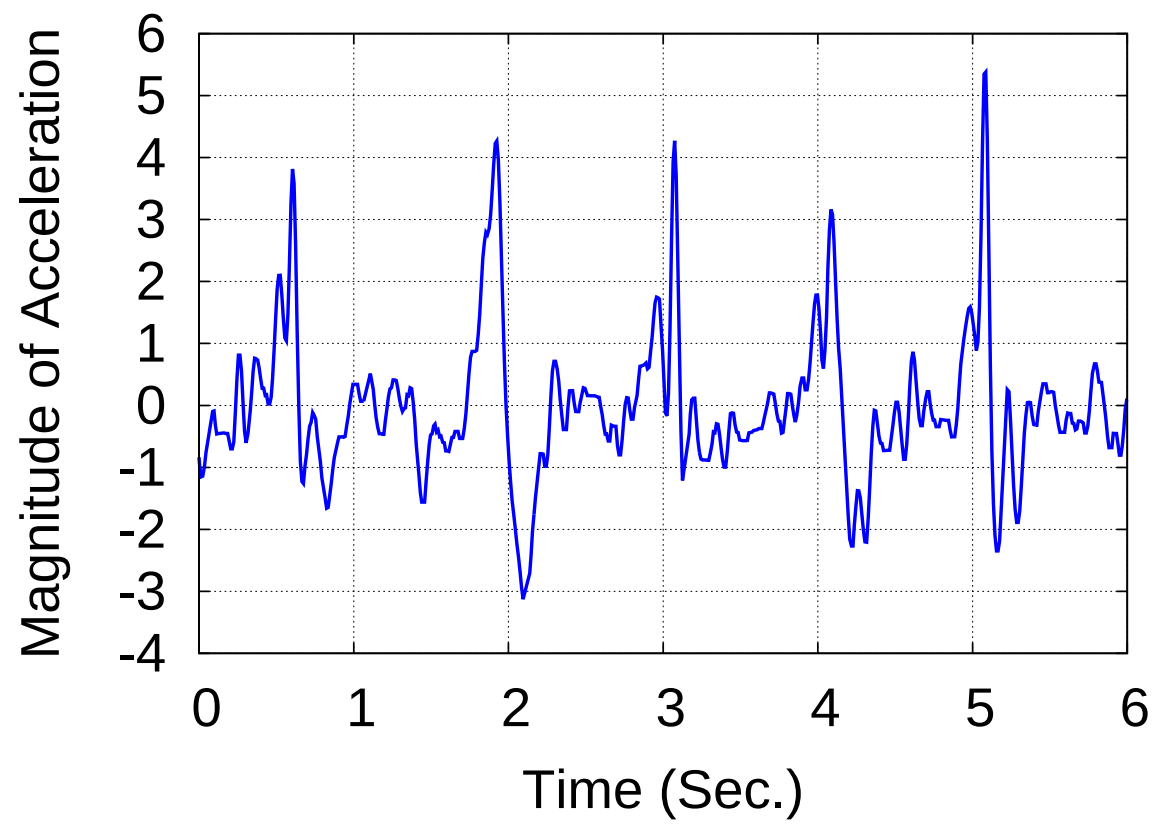

**Figure 3: Pattern of acceleration while walking.**

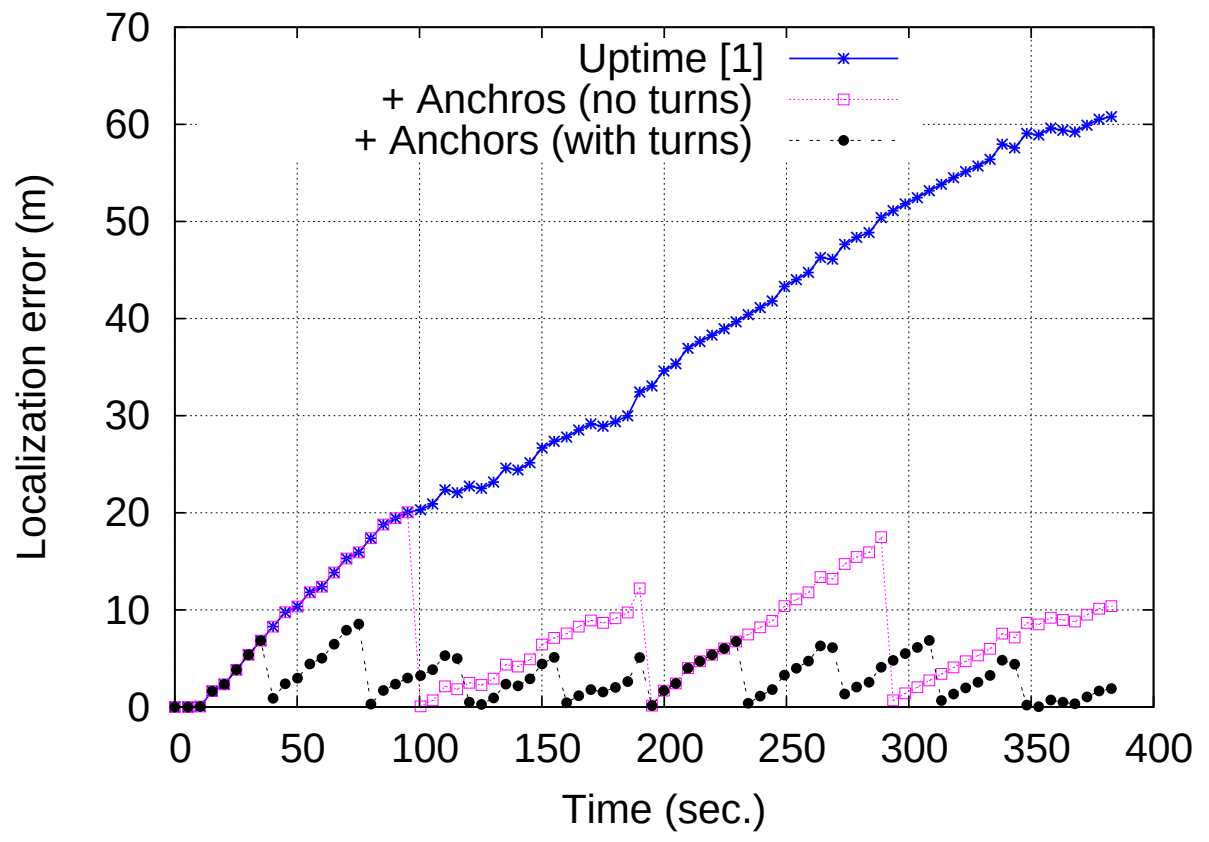

**Figure 4: Effect of using anchor points to reset the error in dead-reckoning.**

user approaches the border of the building.)

## 2.2 Traces Generation Module

The task of this module is to generate accurate motion traces based on the sensors measurements supplied from the data collection module. This module will be running in the cloud to process the raw sensor data from different users.

### 2.2.1 *Background*

Since the main goal is to trace the users inside a building, we cannot depend on the GPS as it requires direct line-of-sight to the satellites. Instead, a ubiquitous localization technique that works with cell phones without infrastructure is required. To address this problem, we rely on a dead-reckoning based approach. In dead-reckoning the current location $(X_k, Y_k)$ is estimated with the help of the previous location $(X_{k-1}, Y_{k-1})$, distance traveled $(S)$, and direction of motion $(\theta)$ since the last estimate as:

$$X_k = X_{k-1} + S * cos(\theta) \qquad (1)$$

$$Y_k = Y_{k-1} + S * sin(\theta) \qquad (2)$$

$\theta$ can be estimated from the magnetometer and/or the gyroscope [21], while the displacement $S$ can be obtained from the accelerometer [1]. The initial position is the last known GPS coordinate, detected by the loss of the GPS signal.

Theoretically the distance traveled can be calculated by integrating acceleration twice with respect to the time. However due to the presence of noise in the accelerometer output, error accumulates rapidly with the time. Another source of error is the presence of a component of acceleration due to the gravity of earth. These factors lead to errors in displacement that will grow cubically with time and can reach 100 meters after one minute of operation even with accurate foot-mounted inertial senors [23]. This error accumulation still exists even if we use the zero velocity update technique [16].

To reduce the accumulation of errors, we **extend** the pedometer-based approach in [1], where the distance traveled is estimated as the sum of the individual step sizes. We apply the step detection algorithm to detect the pattern that the magnitude of acceleration goes through when a step is made. This makes the error in displacement linear in time rather than cubical [1]. Since the pattern of the magnitude of acceleration during walking (Figure 3) is independent from the phone orientation, this makes our approach for displacement estimation independent from the placement of the mobile phone (e.g. in hand, in pocket, etc.). The next section gives the details of our extensions.

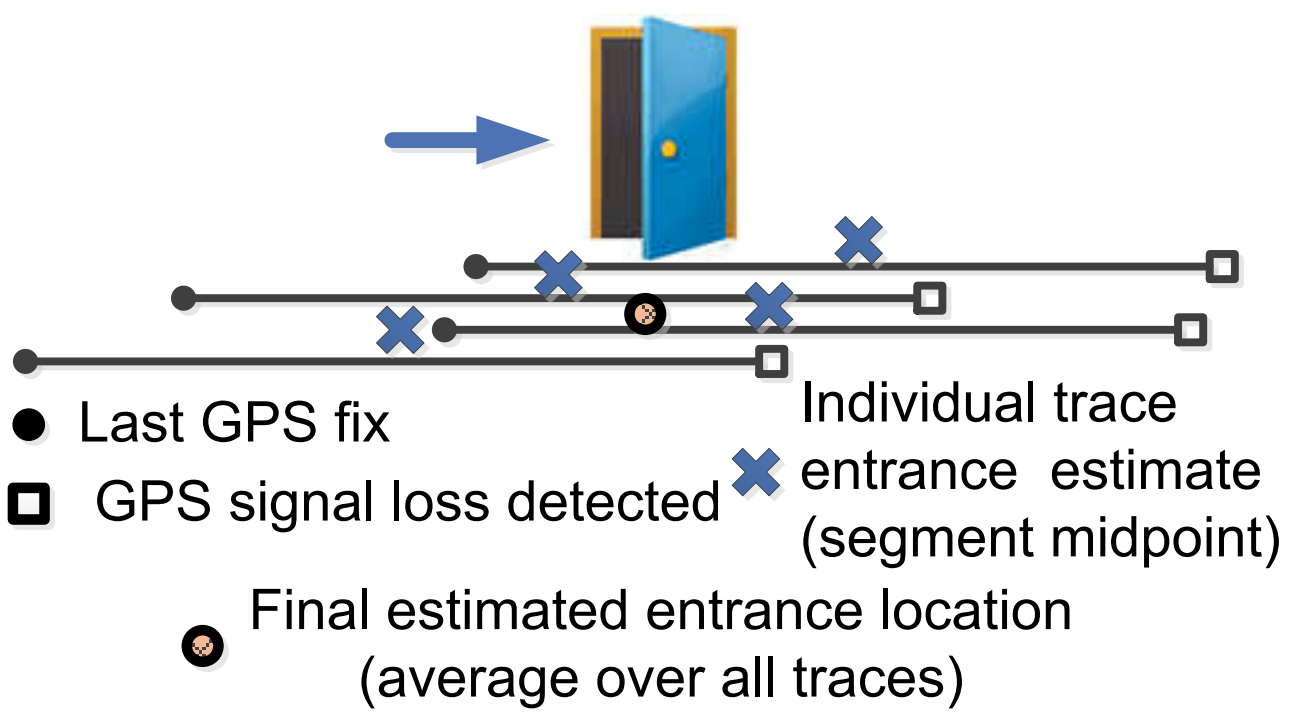

**Figure 5: Estimating the building entrance location using samples from different users. Averaging the midpoint of the segments converges to the actual entrance position as the number of samples increases.**

### 2.2.2 *Proposed anchor-based error resetting*

Even though with a pedometer-based approach for trace generation, there are still errors that lead the generated trace to deviate from the actual motion pattern as time goes by. We believe that the inaccuracy in the traces is due to two main reasons:

1. Error due to the inaccuracy in estimating the trace starting point.
2. Error accumulation of displacement with time, which has been reduced to linear with time using the proposed technique [1].

To further reduce these two sources of error, we introduce the notion of **anchor points**. Anchor points are points in the environment with unique sensor signatures that can be used to reset the trace error when the user hits one of them as shown in Figure 4. In particular, we identify two classes of anchor points: those based on the GPS sensor (building entrances and windows) and those based on inertial sensors (stairs, elevators, escalators, room doors, etc).

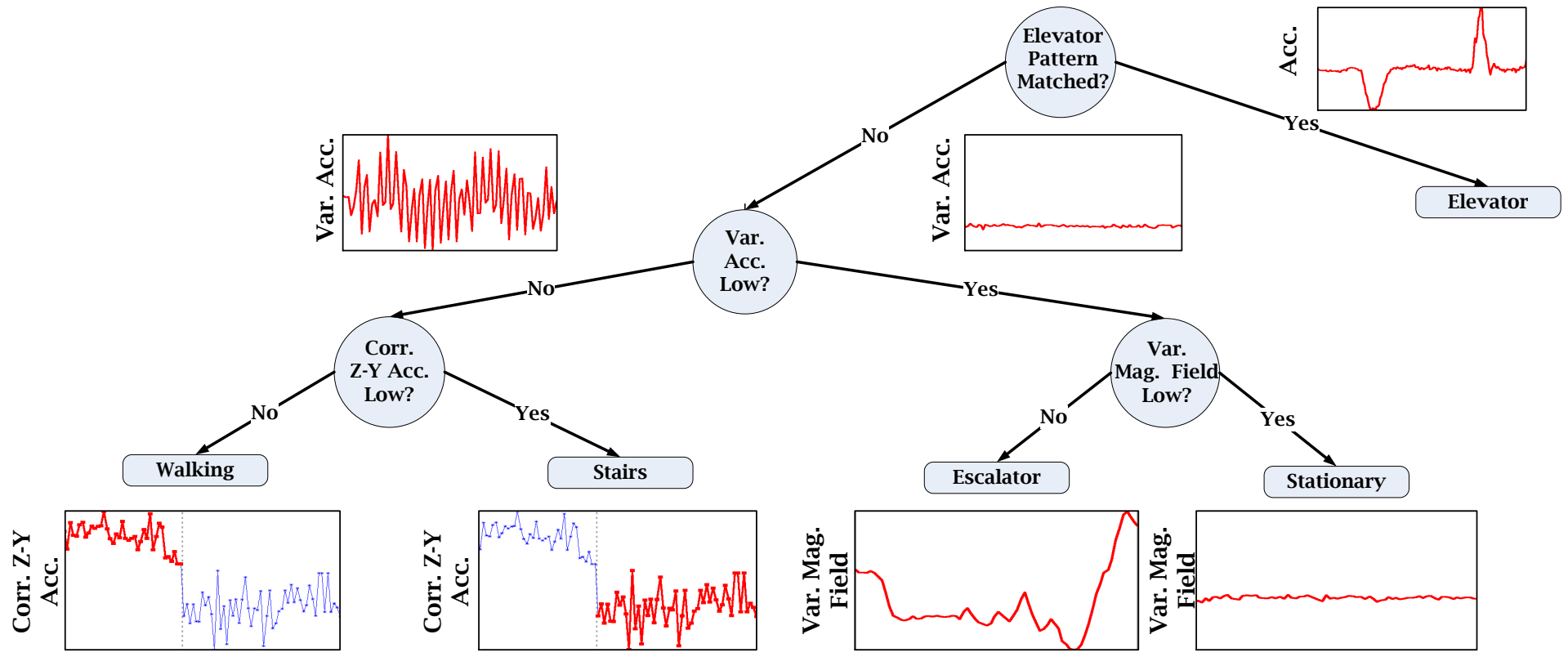

**Figure 6: Classification tree to separate the different inertial-based anchor points.**

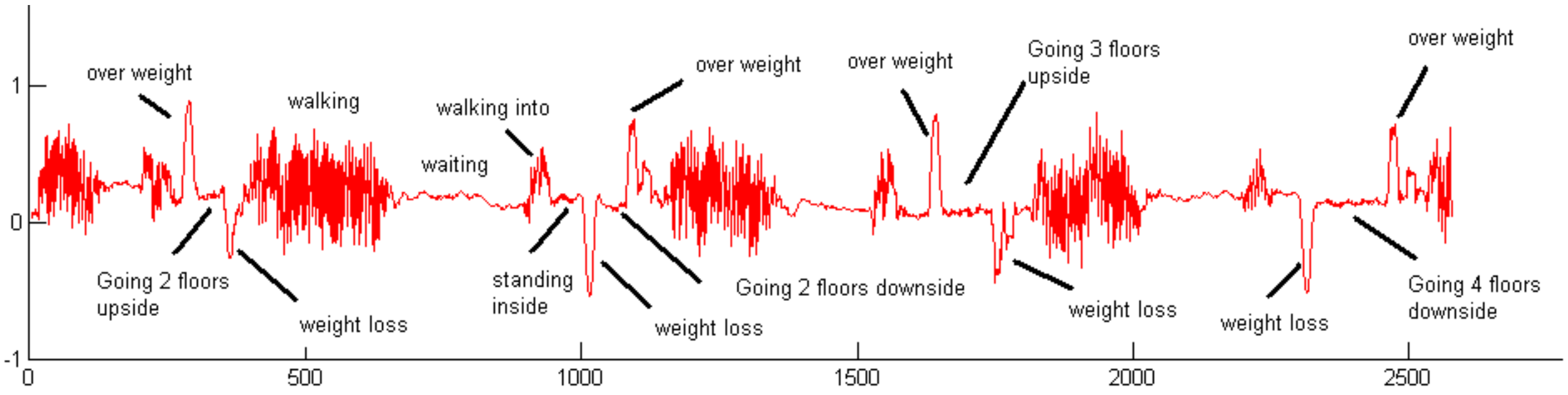

**Figure 7: The acceleration magnitudes of a trace that includes taking an elevator four times. The elevator pattern is unique and can be used to distinguish the direction of movement and number of floors.**

The next two subsections gives the details of identifying these two classes of anchor points.

### 2.2.3 GPS-based Anchor Points

The most observable change to the phone sensors when a user enters a building is the loss of the GPS fix. This can be used to detect the building entrance position. A straight forward approach is to estimate that the user is at the door once the GPS fix is lost. However, this requires the GPS to be always on, to obtain good accuracy, which can kill the phone battery quickly. An alternative energy-efficient approach is to run the GPS sensor at a low duty cycle [27]. The shorter the duty cycle is, the higher the energy savings that can be achieved. However, this comes at an increased error in estimating the door location as the loss of the GPS signal cannot be determined unless the GPS sensor is on. Fortunately, leveraging the large number of traces obtained by our crowdsourcing approach, we can reduce the ambiguity of the building entrance location by statistical techniques.

In particular, the building entrance location is uniformly distributed in the interval between the last obtained GPS position and the first loss of the GPS signal (Figure 5). Using the law of large numbers, the building entrance position can be estimated with high accuracy by averaging a large number of samples as quantified in Section 4.2.1.

Therefore, whenever the loss of the GPS signal is detected, the user position can be reset based on the position of the nearest building entrance/window, enhancing the trace accuracy. This also helps in reducing the error in the trace starting point.

### 2.2.4 Inertial-based Anchor Points

This class of anchor points are based on using the inertial sensors, i.e. the accelerometer, compass (magnetometer), and gyroscope. These sensors have the advantage of being ubiquitously installed on a large class of smart phones, having a low-energy footprint, and being always on during the phone operation (to detect the change of screen orientation). Our focus in this section is on defining a set of rules that enable us to clearly identify elevators, escalators, and stairs as an example of the anchor points that can be identified using the inertial sensors and separating them from other patterns such as normal walking and being stationary. [1]

Figure 6 shows a classification tree for detecting the three classes of interest: elevators, escalators, and stairs. Note that a false positive leads to errors in estimating the location of the anchor point while a false negative leads to missing an opportunity for synchronization. Therefore, high accuracy in detection with low false positive and negative rates are highly desired. We note also that different features can be used to detect the same class accurately, as we show below. This highlights the promise of extracting accurate anchor points.

[1]Note that other virtual anchors, such as turns, can be detected by the inertial sensors in a similar manner.

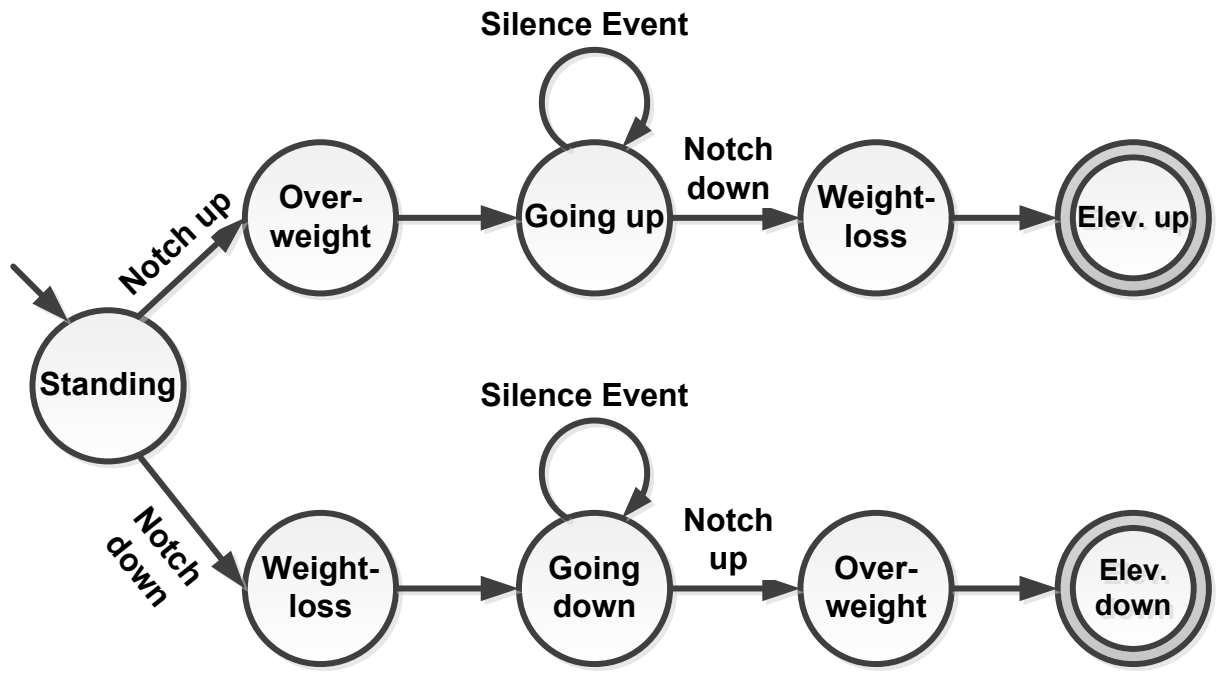

**Figure 8: Finite State Machine to detect the elevator motion pattern.**

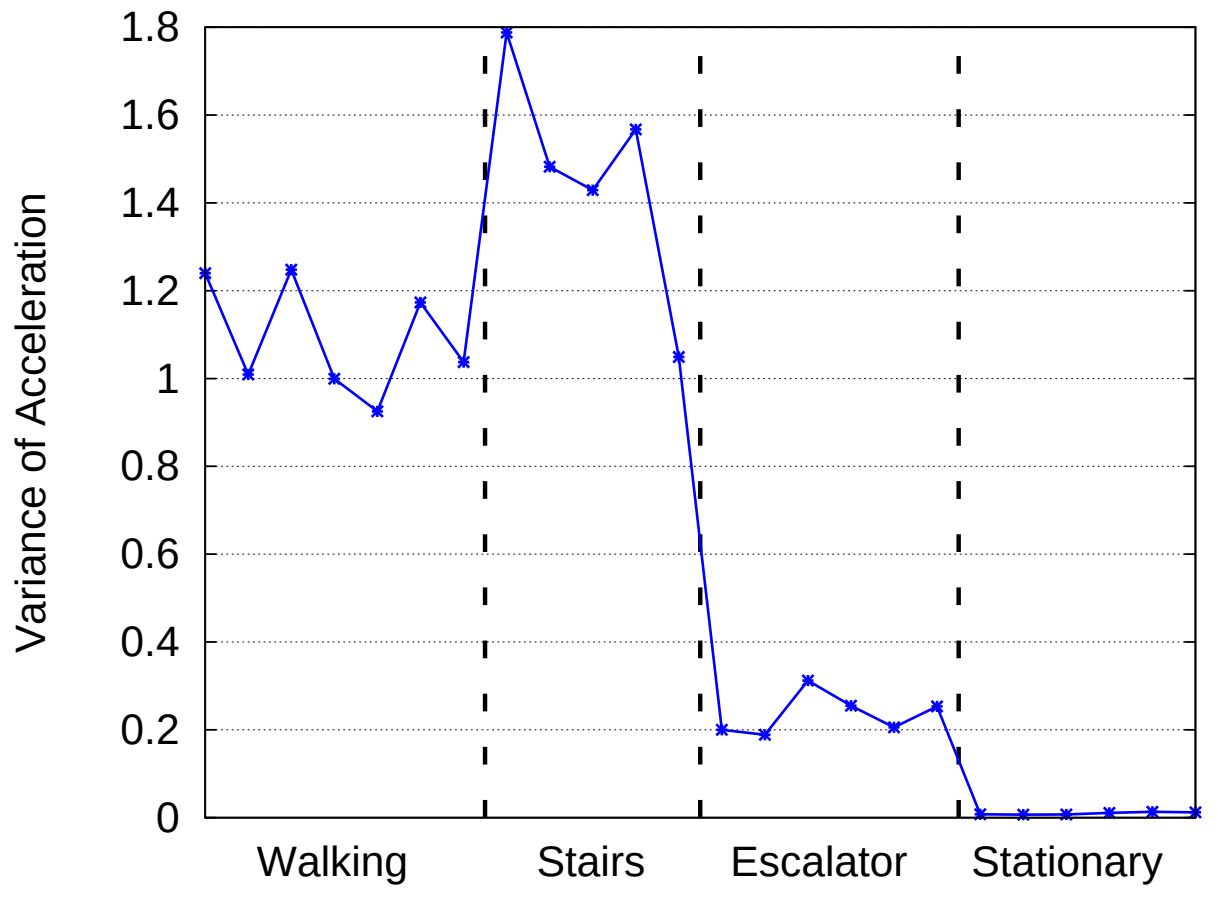

**Figure 9: Variance of acceleration magnitude for different scenarios. The constant speed scenarios (stationary and escalator) have significantly lower variance as compared to the other scenarios.**

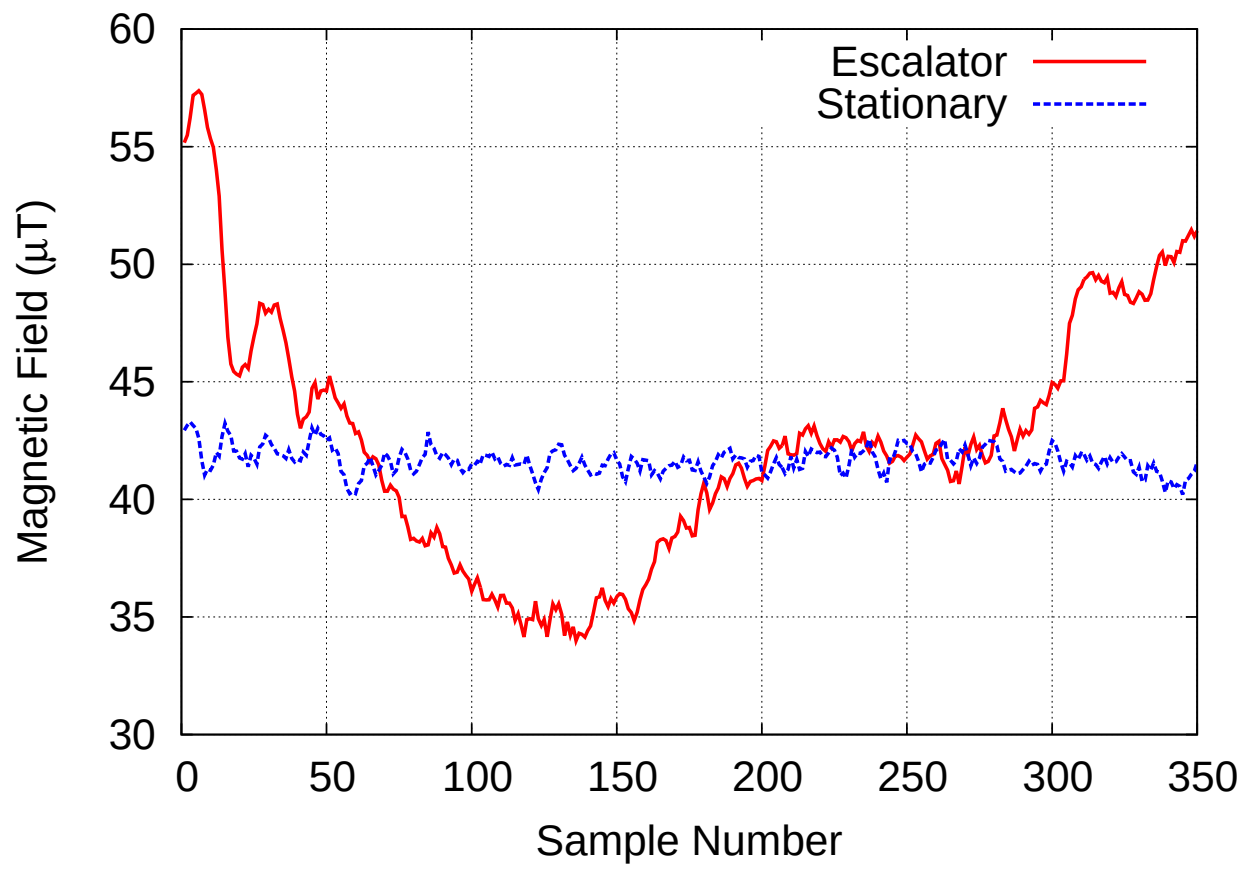

**Figure 10: Magnetic field magnitudes of the escalator and stationary scenarios. The variance of the magnetic field when the user is stationary is much less than the case when she is using an escalator due to the change of location in the case of the escalator and the presence of the powerful motor of the escalator.**

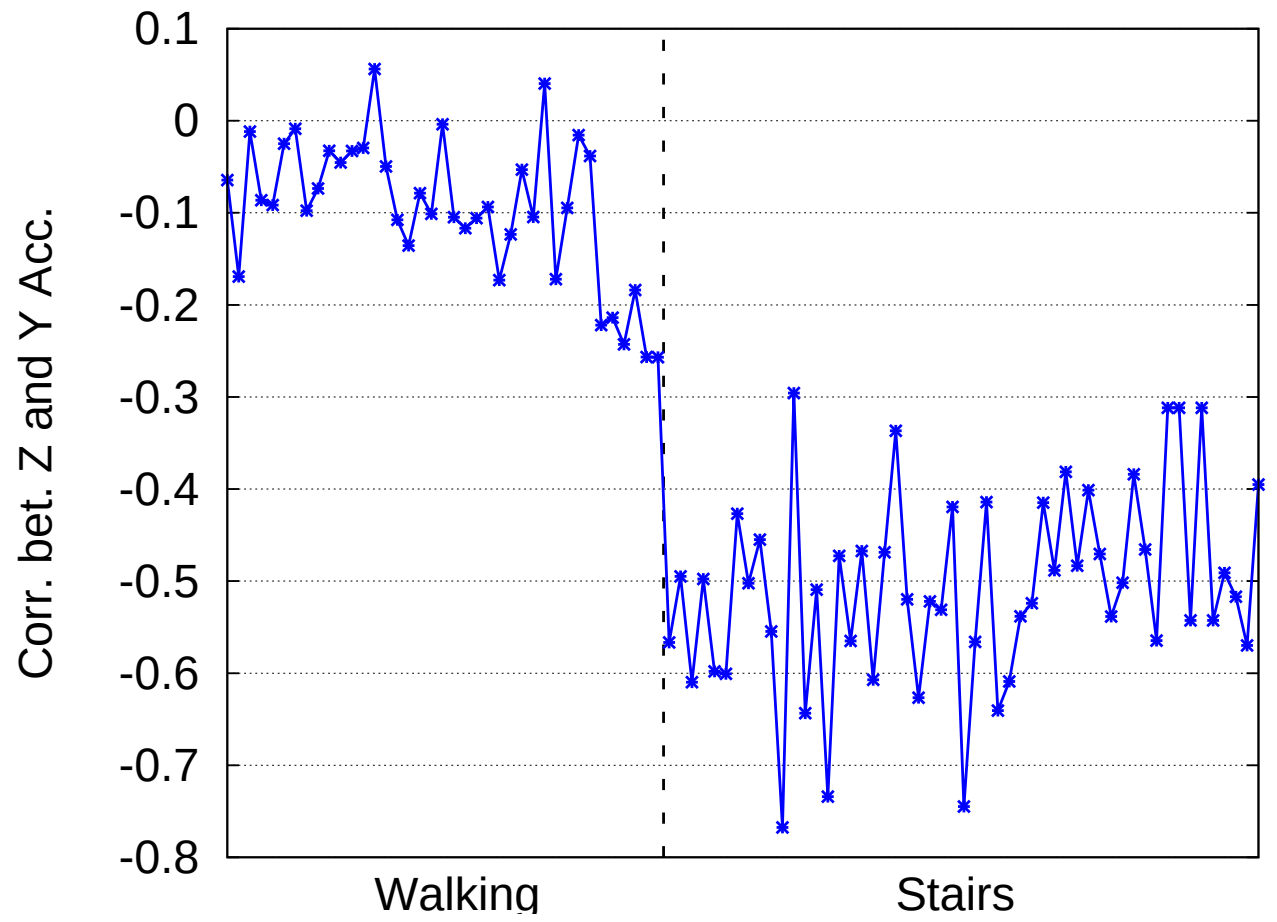

**Figure 11: Correlation between the acceleration values on the different axes for the stairs and walking cases.**

**Elevator:**

The elevator has a unique acceleration pattern that makes it easily distinguishable with high accuracy (Figure 7). A typical elevator usage trace consists of a normal walking period, followed by waiting for the elevator for sometime, walking into the elevator, standing inside, an over-weight/weightloss occurs (depending on the direction of the elevator), then a stationary period which depends on the number of the floors the elevator moved, another weight-loss/over-weight period, and finally a walk-out. To recognize the elevator motion pattern, we developed a Finite State Machine (FSM) that depends on the observed state transitions (Figure 8). The detected direction of motion (based on the order of the weight-loss over-weight events) and the number of floors traveled (based on the time or displacement during the inside-elevator period), can be used to further enhance the accuracy.

**Escalator:**

Once the elevator has been separated, the key observation that distinguishes the constant speed scenarios (escalator/stationarity) from the dynamic scenarios (stairs/walking) is that users do not move their legs in the constant speed scenarios. Moving legs has a significant effect on the variance of the acceleration pattern (Figure 9).

To further separate the escalator from stationarity, we found that the variance of the magnetic field when the user is stationary is much less than the case when she is using an escalator (Figure 10). We believe that this is due to the change of location in the case of the escalator and the presence of the powerful motor of the escalator.

**Stairs:**

Once the dynamic scenarios have been separated based on the variance of acceleration, what remains is to differentiate between the stair and walking cases. Figure 11 shows that the correlation between the acceleration in the $Y$ and $Z$ axes can be a good measure to separate the two cases. The intuition is that when the user is using the stairs, her speed increases or decreases based on whether the gravity is helping her or not. This creates a higher correlation between the acceleration in the direction of motion and direction of gravity as compared to walking.

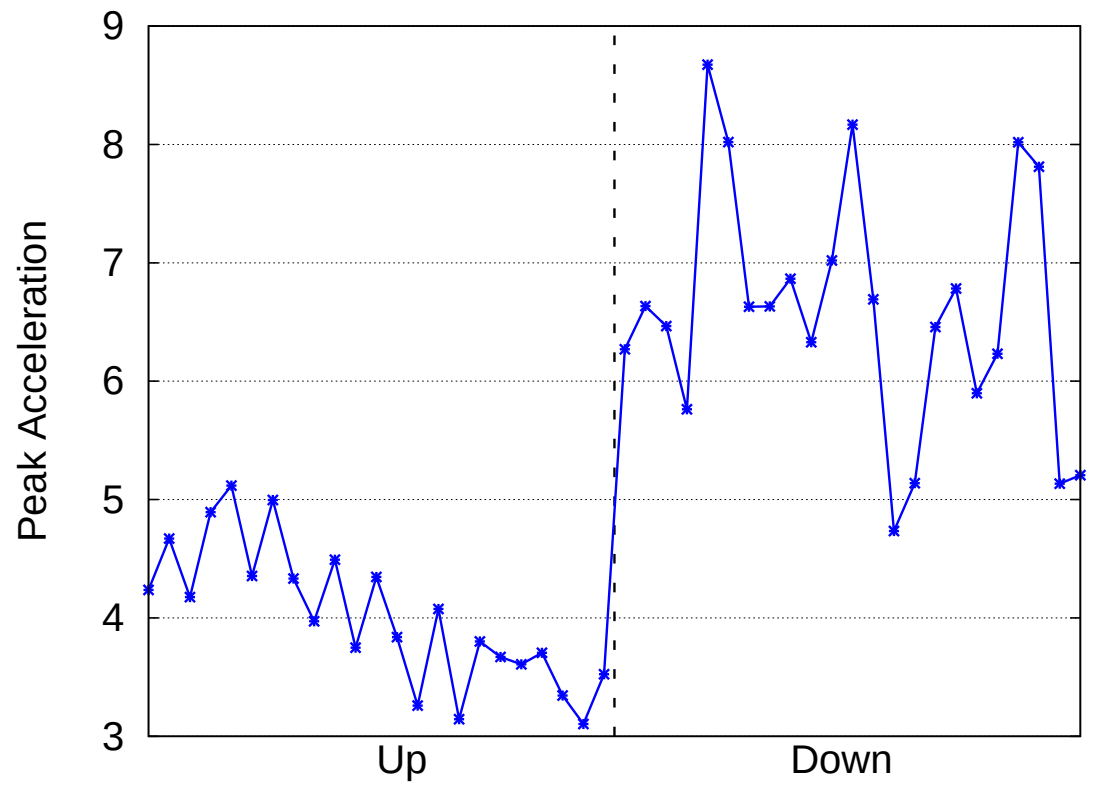

**Figure 12: Peek of acceleration magnitude for the different stair climbing directions.**

Furthermore, our measurements show that climbing down stairs exhibit a higher motion intensity than climbing up (as the gravity is helping the user in the former case). This is reflected in different features. For example, Figure 12 shows that the peak of the acceleration magnitude can be used to differentiate between the stairs up/down case.

#### *2.2.5 Estimating the Location of Inertial-based Anchor Points*

Similar to the case of GPS-based anchor points, to estimate the location of an inertial-based anchor point we use the average location of all position estimates for users coming by this anchor point. Once the location of an anchor point is found, it is added to the generated floorplan and also used to enhance the newly generated traces.

Note that the location of the anchor points can be estimated with a few number of traces and the accuracy increases with more traces (as quantified in Section 4). In addition, the WiFi signal and AP MAC addresses can be used to distinguish between the different anchor points of the same kind. This is especially needed in the case of dense anchors from the same type, such as using turns as virtual anchor points.

## 3. FLOORPLAN ESTIMATION MODULE

Once accurate motion traces are collected from different users, the goal of this module is to estimate the building floorplan. There are two levels of details that can be obtained: (1) the overall shape and (2) the room-corridors details. We start by giving the details of both modules and end the section by a discussion of other higher level semantic information that can be obtained.

### 3.1 Overall Floorplan Shape

This level of detail provides a black and white occupancy map of the building. In particular, areas where users move represent walkable area (black) and areas free of users' traces represent blocked area (white).

In order to automatically estimate the overall floorplan shape, we represent each ***user step*** by a point. The goal is to estimate the best shape that represents the **point cloud** generated from all collected traces (Figure 13(b)).

We found that alpha shapes is a general tool that can capture the building shape with high accuracy. An alpha shape ($\alpha$-shape) is a family of piecewise linear simple curves in the Euclidean plane associated with the shape of a finite set of points [7]. An edge of the $\alpha$-shape is drawn between two points in the set if there exists a generalized disk of radius $1/\alpha$ containing the entire point set and which has the property that the two points lie on its **boundary**. Note that for $\alpha < 0$, this is equivalent to drawing an edge between two points of the set if there exists a generalized disk of radius $1/\alpha$ which has the property that the two points lie on its boundary and does not contain any of the remaining points. The $\alpha$-shape is a generalization of the concept of the convex hull (for $\alpha = 0$).

Figure 13(d) shows the $\alpha$-shape of the point cloud in Figure 13(b). Note that the convex hull of the same point cloud cannot capture the concavity and holes in the floorplan shape (Figure 13(c)).

### 3.2 Detailed Floorplan

To further obtain more details about the building internals, we apply a number of processing steps on the collected traces to discover the distinct rooms, corridors, and room doors. These include traces segmentation and filtering, segments classification into rooms and corridors, segments clustering to obtain rooms boundaries, estimating the doors positions, and final shaping and labeling. We give the details of these modules in the following subsections.

#### *3.2.1 Traces segmentation and filtering*

The first step in our approach is to break the continuous motion traces into segments. Segments are straight parts of the trace that are separated by either turns or pauses (periods of inactivities). Although there are advanced techniques for segmenting trajectories of outdoor traces, e.g. [4], based on our experiments we found that a simple segmentation algorithm based on the heading change is sufficient for segmenting our traces. In particular, consecutive segments are separated by significant changes in the direction of motion (we have chosen the threshold to be $45°$). The intuition is that a segment will be inside the same area (corridor/ room/ hall). Figure 14(a) shows how a sample trace has been broken into 10 segments (each segment shown in a different color).

Finally, we filter the segments by excluding short segments in terms of both time and/or distance as we found that those segments are not descriptive.

#### *3.2.2 Segments classification*

The goal of this module is to identify the type of each segment as one of two categories: corridors or rooms. Once identified, the clustering and shaping modules described in the next section determine the rooms and corridors areas. We use a standard tree-based classifier using the following features:

- **Average time spent per step in the segment:** This feature represents the average time spent between individual steps in the segment. The intuition is that, typically, the user walks faster through corridors than rooms.
- **Segment length:** Since we start a segment at each significant

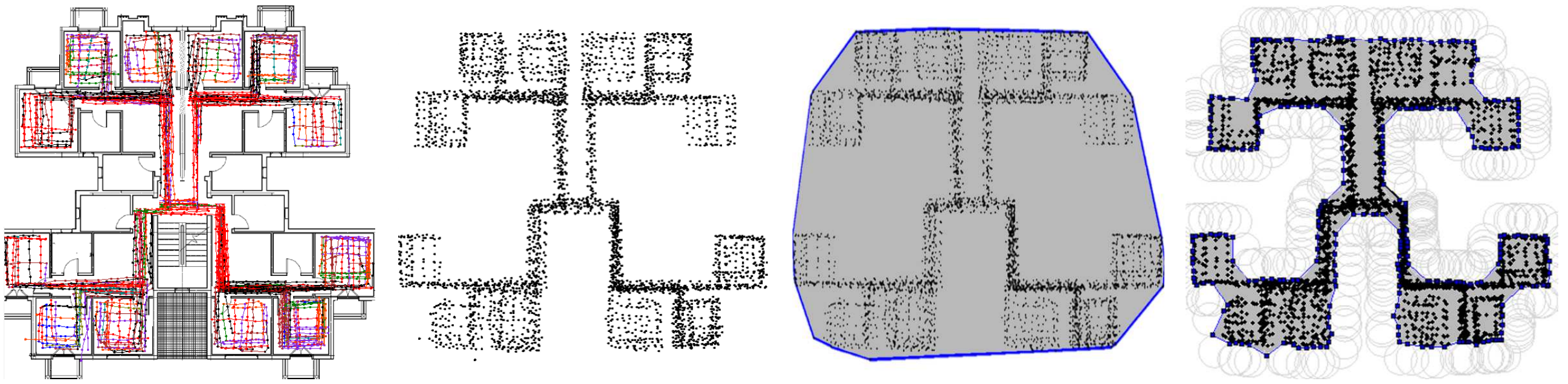

(a) Collected traces

(b) Point cloud

(c) Shape of the building using the convex hull

(d) Shape of the building using $\alpha$-shapes

**Figure 13: Construction of the overall building floorplan from multiple motion traces. The grey area in subfigure (d) represents the estimated floorplan shape. The convex hull of the same point cloud (subfigure (c)) cannot capture the concavity and holes in the floorplan shape.**

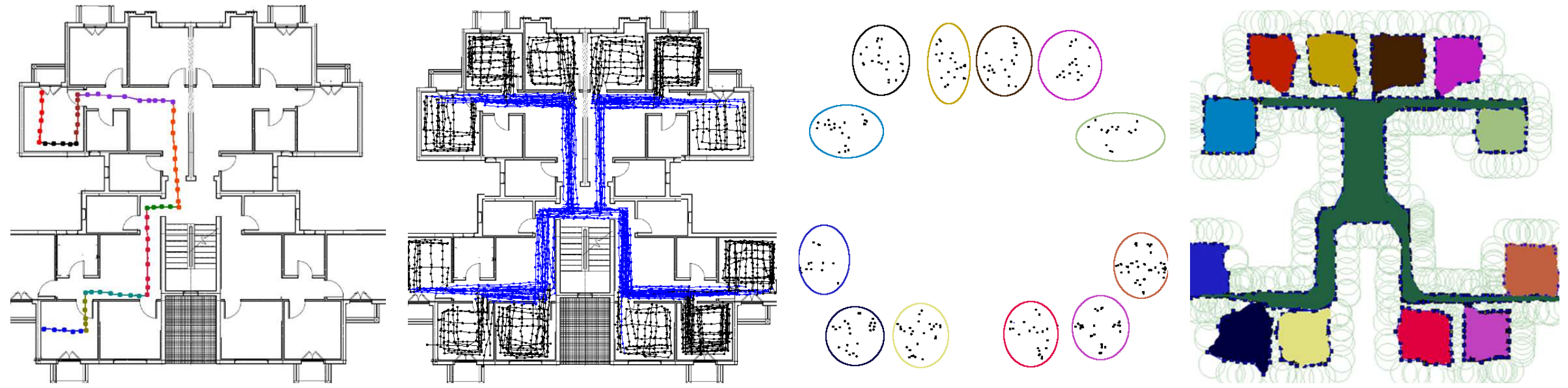

(a) Breaking a trace into segments

(b) Classified segments (blue corridors, black rooms)

(c) Room clusters based on the center points of segments

(d) Detailed estimated floorplan

**Figure 14: Construction of a detailed floorplan using multiple motion traces.**

change in direction, this feature captures the intuition that the segments in corridors should be longer than rooms.

- **Neighbor traces density:** The intuition here is that the segments in the corridors are more dense (as more users use them) than the segments in rooms (as shown in the point cloud in Figure 13(b)).

The result of classification is shown in Figure 14(b). Segments that are classified as "Corridors" are drawn in blue, whereas those classified as "Rooms" are shown in black.

#### 3.2.3 *Segments clustering*

Once we identified the type of each segment (i.e. room or corridor), we apply a clustering algorithm on all segments of type "room" to find the number of rooms, their boundaries, and where they exist. We use a density-based clustering algorithm (DBSCAN) to group segments that lie close to each other into one cluster. To prevent segments from adjacent rooms to be grouped together and reduce the effect of the traces noise that may cross the walls between rooms, we use the center point of each segment for the clustering (rather than all the points in the segment). The similarity measure used for clustering is the distance between the location of **center points** and the similarity between the measured WiFi signals at these points. Figure 14(c) shows the clusters generated using the segments center points.

#### 3.2.4 *Shaping*

To estimate the shape of rooms, we calculate the $\alpha$-shape of the points corresponding to all the segments that belong to each room separately (as generated by the clustering module). Similarly, to obtain the corridor shape, we find the $\alpha$-shape of the complete corridors point set. The final estimated floorplan is shown in Figure 14(d) where different rooms are shown in different colors.

Note that further smoothing can be applied to the obtained rooms and corridors, e.g. to make them rectangular. However, this may not work for general shapes such as concave corridors.

#### 3.2.5 *Estimating doors positions*

To estimate the locations of the room-doors, we extract all the intersection points of two segments; one of which is of a *corridor* type while the other is of a *room* type. The distribution of those points of intersection is shown in Figure 15(a). We apply a spatial clustering algorithm (DBSCAN) on these points based on the Euclidean distance between points as a similarity measure. Each cluster corresponds to a door whose centroid is taken as the estimated door location. Figure 15(b) shows the estimated locations of doors.

### 3.3 Discussion

Higher levels of semantic can be attached to the estimated floor plan. These include identifying the room (shop in case of a mall) type (e.g. a restaurant versus a book shop), shop brand (e.g. KFC vs. Starbucks), and/or room owner. Different approaches can be investigated for these semantic labels including: (1) using the different sensors on the phone (e.g. camera and mic) to fingerprint some locations, (2) integration with social network information and individual visiting patterns to a certain location, (3) asking users in unknown or ambiguous locations to identify the location label in a form of a game (similar to digitizing books using reCAPTCHA), (4) using automatic group discovery, e.g. using bluetooth co-location traces [15], and associating their pattern with their locations for identifying restaurants, classrooms, meeting rooms, and similar areas, and/or (5) using external feedback, e.g. through examining the billing confirmation SMSs that users receive upon making payments using the credit card to identify the location [19].

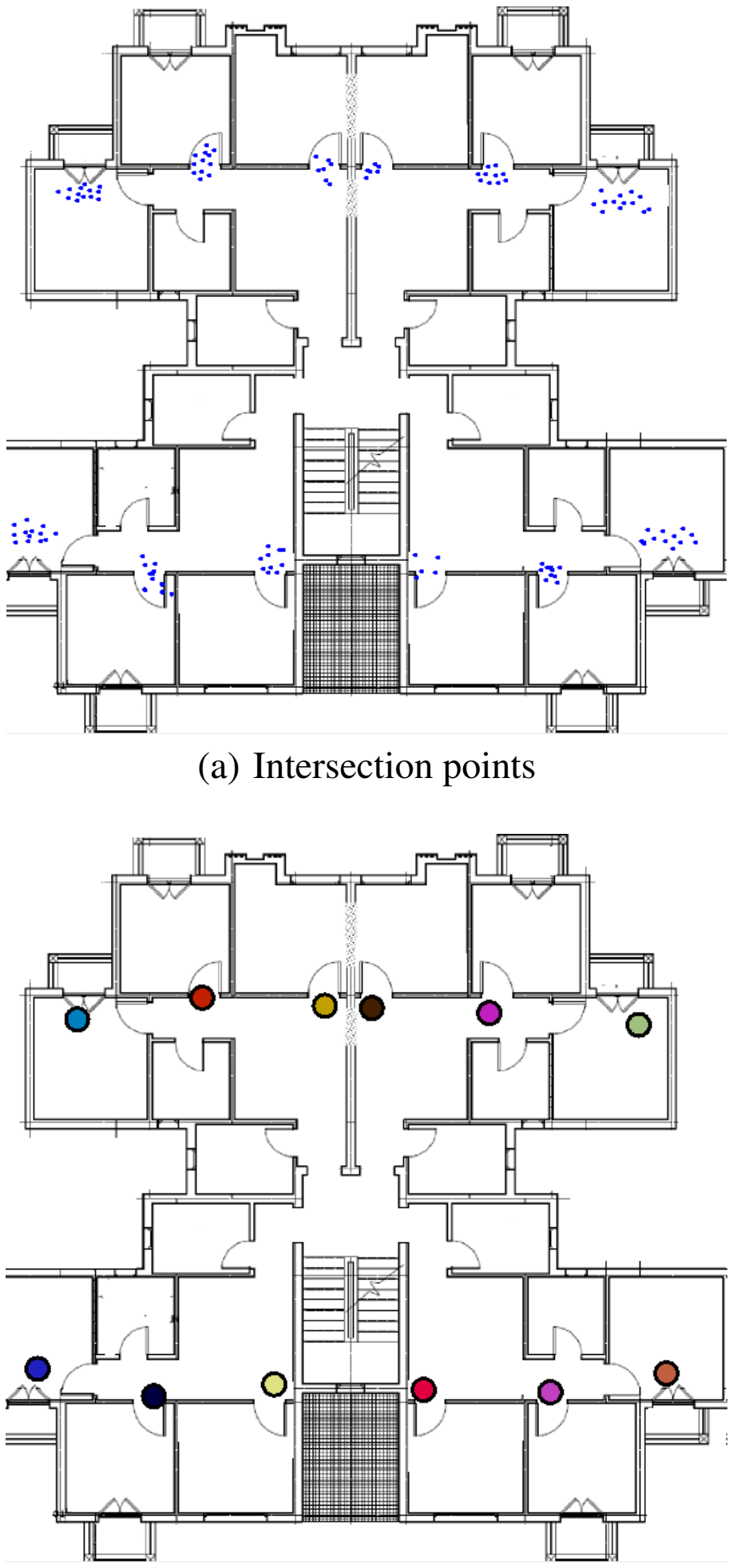

(a) Intersection points

(b) Estimated doors locations

Figure 15: Estimating the locations of rooms doors.

# 4. EVALUATION

In this section, we evaluate the performance of the *CrowdInside* system in two typical testbeds. We start by describing the testbeds followed by evaluating the performance of the anchor points estimation accuracy, trace generation accuracy, and the floorplan estimation accuracy.

## 4.1 Testbeds

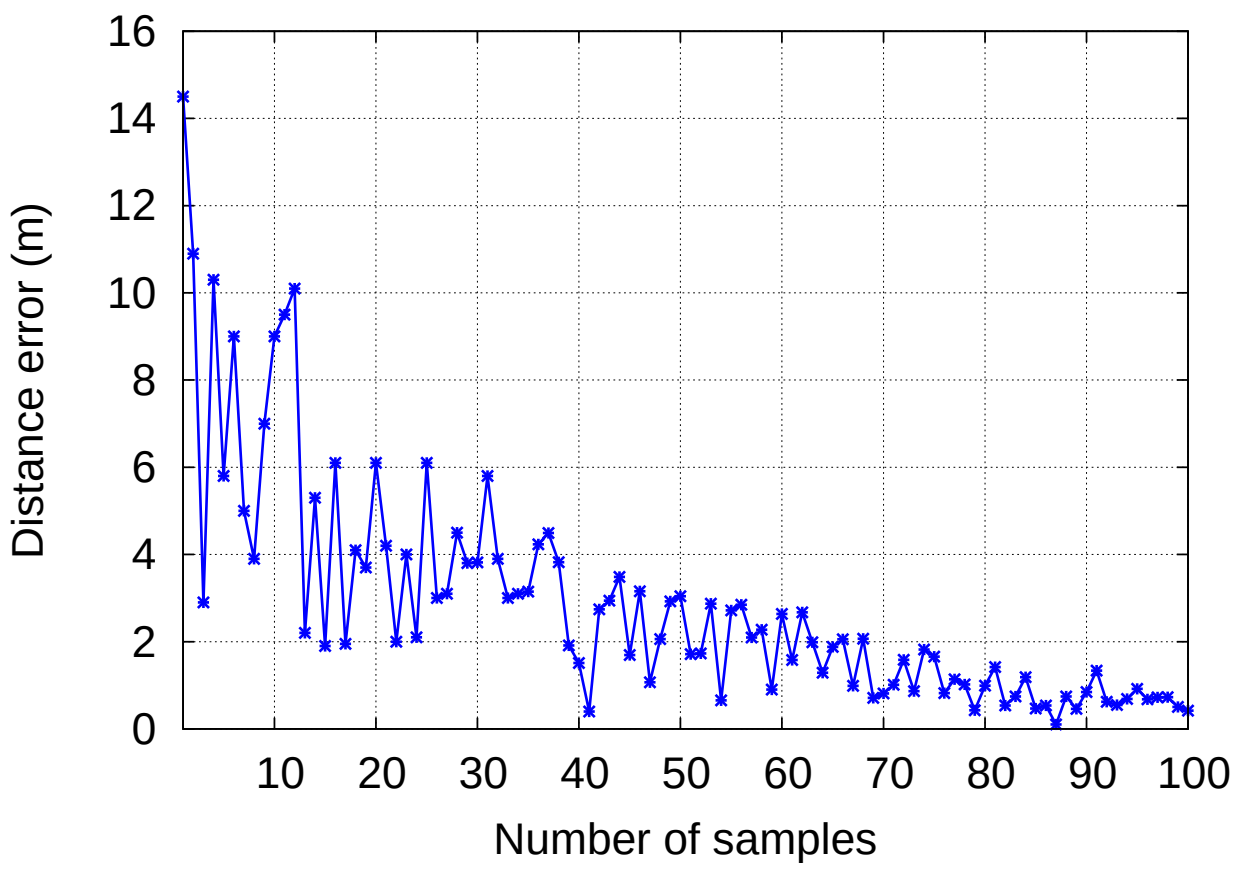

**Figure 16: Accuracy of the building entrance position estimation against the number of samples.**

We implemented our system on different Android phones (Samsung Nexus S, Nexus One, Galaxy Ace and Galaxy Tab). We carried out our experiments in two testbeds: a shopping mall with plenty of stairs/elevators/ escalators and a building in our campus with an approximately $448m^2$ area. The first testbed is used to evaluate the accuracy of trace generation and anchor-based error resetting. The second testbed is used for evaluating the floorplan construction as we had access to most of the rooms. Around 100 traces were collected with the help of four volunteers covering 12 rooms along with all corridor areas.

For querying the sensors, we used the lowest sampling rate (UI_Delay), which is the rate used to detect the screen orientation change. This has the advantage of using no extra energy for the trace generation over the normal energy for querying the inertial sensors.

## 4.2 Anchor Points Estimation Accuracy

### 4.2.1 GPS-based anchor points

We collected 100 random traces, with a two minutes duty cycle, each of them starts outdoors and ends inside the building. The actual location of the building entrance was recorded manually as a ground truth. Figure 16 shows the accuracy of estimating the entrance position. The figure shows that as the number of samples increases, the accuracy increases significantly, reaching less than 1m error with only 100 samples.

Figure 17 shows the effect of changing the duty cycle on the number of samples required to achieve an error less than a specific accuracy with a 95% confidence interval. The figure shows that even with a very low duty cycle of 6 minutes (turning on the GPS once every 6 minutes), as low as 1200 samples are required to obtain an error of less than one meter with a 95% confidence. Moreover, these samples are amortized over the number of users who use a certain building, highlighting the ability of *CrowdInside* to quickly, accurately, and efficiently (in terms of energy) estimate the building entrance location.

### 4.2.2 Inertial-based anchor points

To evaluate the inertial-based seed anchor points detection accuracy, we collected 152 traces covering the three different classes (elevators, escalators, and stairs) along with walking, and stationarity. Table 1 shows the confusion matrix for the different classes.

Table 1: Confusion matrix for classifying different inertial anchors

| | Elevator | Stationary | Escalator | Walking | Stairs | FP | FN | Total |
|---|---|---|---|---|---|---|---|---|
| **Elevator** | 24 | 0 | 0 | 0 | 0 | 0% | 0% | 24 |
| **Stationary** | 0 | 31 | 1 | 0 | 0 | 0% | 3.1% | 32 |
| **Escalator** | 0 | 0 | 20 | 0 | 0 | 0.65% | 0% | 20 |
| **Walking** | 0 | 0 | 0 | 30 | 0 | 0.65% | 0% | 30 |
| **Stairs** | 0 | 0 | 0 | 1 | 45 | 0% | 2% | 46 |
| **Overall** | | | | | | 0.2% | 1.3% | 152 |

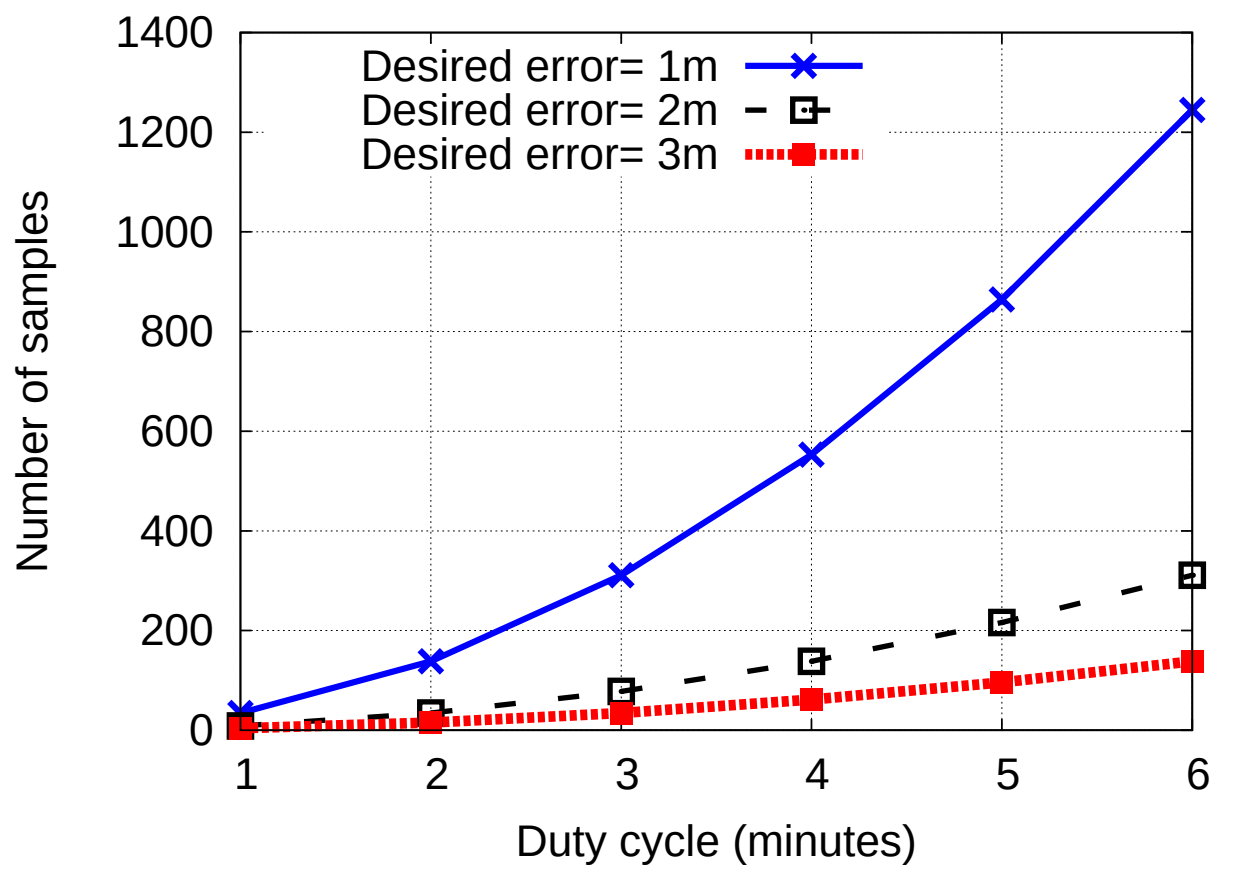

**Figure 17: Building entrance location estimation as a function of duty cycle. The y-axis repents the number of samples required to achieve a specific accuracy with a 95% confidence interval.**

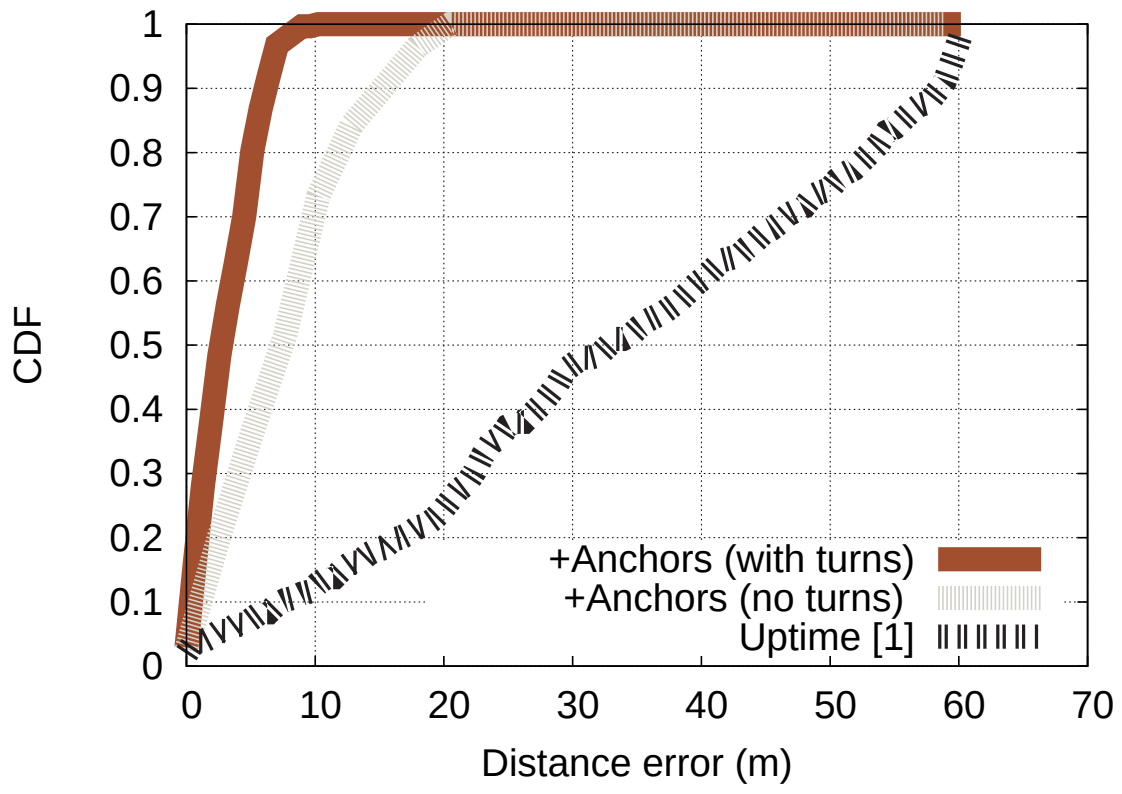

**Figure 18: CDF of displacement error using the proposed anchor-based error resetting techniques.**

The table shows that the different classes can be separated with high accuracy (less than 0.2% FP rate and 1.3% FN rate) using our classification tree in Figure 6.

### *4.2.3 Trace generation accuracy*

Figure 18 shows the CDF of displacement error using the proposed anchor-based error resetting techniques as compared to the state-of-the-art [1]. The figure shows that using anchor points, we can achieve significant enhancement in accuracy, up to 12 times, as compared to the state-of-the-art.

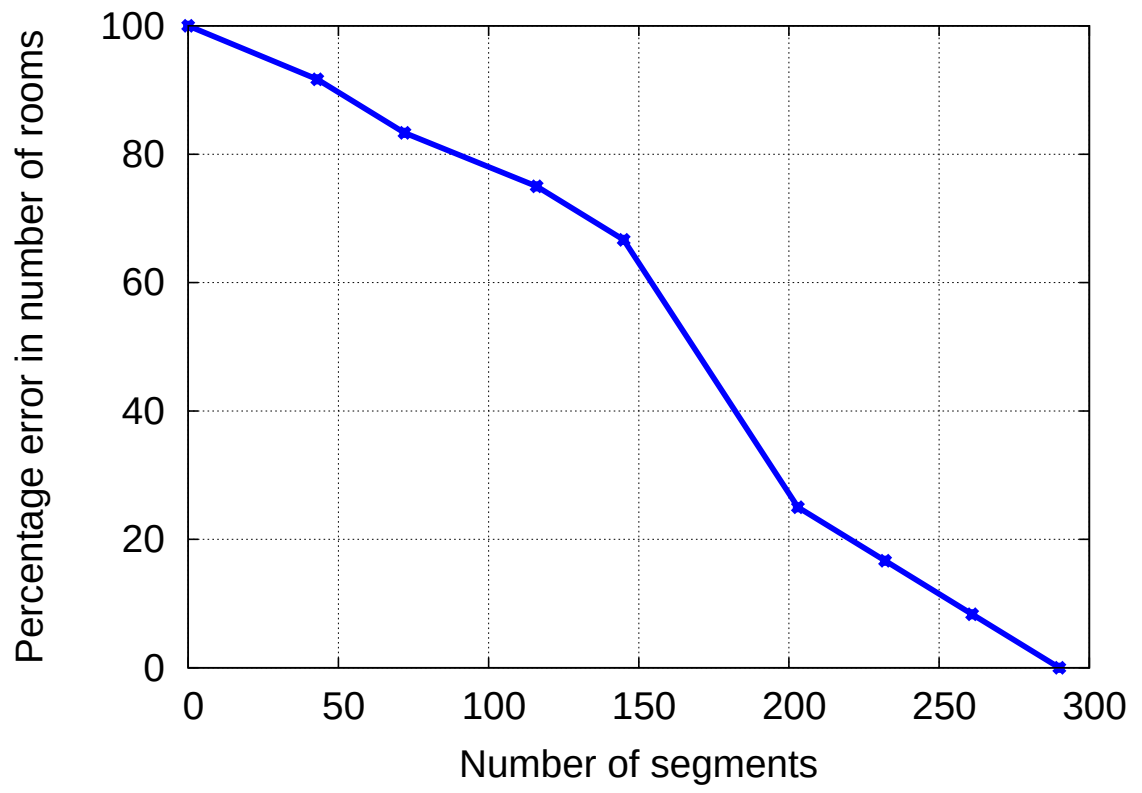

**Figure 19: Effect of number of collected segments on percentage error in number of rooms.**

## 4.3 Floorplan Estimation Accuracy

Figure 19 shows the effect of increasing the number of traces (segments) on the accuracy of the generated floorplan. Figure 14(d) shows the generated floorplan for the different number of segments. The figures show that as few as 290 segments are required to obtain the full floorplan shape. The break at about 150 segments is due to the traces starting to cover all rooms in the floorplan.

# 5. RELATED WORK

Many systems over the years have tackled the indoor localization problem including infrared [22, 2], ultrasonic [18], computer vision [13], physical contact [17], and radio frequency (RF) [25, 26, 14, 24, 10, 12, 11] based systems. All these systems are usually deployed in limited areas and assume the existence of building floorplans. In addition, RF-based techniques usually require the construction of radio fingerprints, which is both time and labor intensive.

The introduction of inertial sensors to mobile phones (e.g. accelerometers, magnetometers) offers an opportunity for performing user positioning and tracking using dead-reckoning in a ubiquitous manner and without any prior setup cost [1]. However, since dead-reckoning suffers from accumulation of error, step-counting techniques, e.g. [1, 21], have been proposed to reduce the dead-reckoning error by counting the number of steps rather than integrating the acceleration. In our work, we extend such systems to leverage explicit common environment anchor points, such as doors, elevators and stairs, to reset the errors.

Recently, the idea of relying on user input, i.e. crowd-sourcing, to create a location database has been proposed for both outdoor and indoor localization, e.g. [8, 3, 5]. Such systems focus on con-structing the RF fingerprint database, but not on constructing the floorplan. In addition, some of them require manual user input, which may not be convenient to the user.

Simultaneous Localization and Mapping (SLAM) [6] is a well known technique in the mobile robotics domain which is concerned with solving the problem of localizing a mobile robot moving in an unknown environment while simultaneously building a map of the surrounding area. Typically SLAM employs the robot odometry and high-end laser range sensors, ultrasonic, or computer vision to build maps of the environment. Mapping in SLAM terminology refers to identifying unique signatures in the environment, and not the building map. The types of sensors used by traditional SLAM limits it from being implemented using commodity mobile phones. In addition, SLAM does not fuse the data from multiple robots nor target construction of floorplans or higher semantics. SmartSLAM [20] presents a modified SLAM algorithm for smart phones that employs a pedestrian tracking system using inertial sensors as a motion model along with Wi-Fi signals as an observation model. However, the maps generated by SmartSLAM describes only the corridor layout of the building with no information about the number of rooms, their shapes and locations.

A statistical method for 3D roof reconstruction from laser scan point clouds is described in [9]. However, this method depends on employing airborne laser scanners for generating fine-grained point clouds and is only concerned of roof shapes construction.

In summary, *CrowdInside* is unique in leveraging anchor points to enhance traces accuracy. In addition it allows for automatically detecting the floorplan outline and the floorplan detailed shape.

# 6. CONCLUSION

In this paper we presented the *CrowdInside* system for the automatic construction of indoor floorplans. Our approach is completely autonomous and depends only on the data collected from users moving naturally inside the buildings. We presented a method for enhancing the dead-reckoning accuracy by using unique anchor points which are found in typical indoor spaces for error resetting. Based on the accurate user traces, we described approaches for detecting both the floorplan layout and a more detailed floorplan with rooms, corridors, and doors identified. Neither the generation of traces nor the floorplan estimation require special infrastructure nor prerequisite details about the buildings layout.

We implemented our system using commodity mobile phones running the Android operating system and evaluated it in two testbeds. Our results show that we can detect the anchor points accurately with 0.2% FP rate and 1.3% FN rate. In addition, the proposed error resetting technique leads to more than 12 times enhancement in the median distance error. Moreover, the detailed floorplan can be estimated with as few as 290 segments. These segments are amortized over the number of users using the building.

Currently, we are expanding *CrowdInside* in multiple directions including inferring higher level semantic information, such as rooms types and owners, energy-efficiency aspects, user incentives, among others.

## Acknowledgments

The authors would like to thank Ahmed Elgohary for his help in the experiments. This work is supported in part by a Google Research Award and a TWAS-AAS-Microsoft award.